\documentclass{aa}
\usepackage{graphicx,epsfig,amssymb}

\newcommand{\fgas}{f_{\rm gas}}

\newcommand{\Ho}{H_{\rm o}}
\newcommand{\OmM}{\Omega_{\rm M}}

\newcommand{\OmL}{\Omega_\Lambda}
\newcommand{\Sei}{\sigma_8}
\newcommand{\Da}{D_{\rm ang}}
\newcommand{\da}{d_{\rm ang}}
\newcommand{\npair}{n_{\rm pair}}
\newcommand{\Dtheta}{\Delta\theta}

\begin{document}
   \title{Cosmological constraints from a 2D SZ catalog}

   \subtitle{}

\titlerunning{Constraints from a 2D SZ catalog}

   \author{S. Mei
          \inst{1,2}
          \and
          J.G. Bartlett \inst{3}
          }

   \offprints{S. Mei}

   \institute{ Institut d'Astrophysique Spatiale, UMR-8617, Universit\'e Paris-Sud, B\^atiment 121, F-91405 Orsay, France\\ 
     \and
          Johns Hopkins University, 3400 N. Charles Street, 21218 Baltimore, MD, USA\\
          \email{smei@pha.jhu.edu}
          \and
             APC -- Universit{\'e} Paris 7, 11, place Marcelin Berthelot, 75005 Paris, France\\
             \email{bartlett@cdf.in2p3.fr}
             }

   \date{Received: 9 October 2003/Accepted 7 May 2004}

\abstract{We perform a Fisher matrix analysis to quantify
cosmological constraints obtainable from a 2--dimensional
Sunyaev--Zel'dovich (SZ) cluster catalog using the counts and the
angular correlation function.  Three kinds of SZ survey are
considered: the almost all--sky Planck survey and two deeper
ground--based surveys, one with 10\% sky coverage, the other one with
a coverage of 250 square degrees.  With the counts and angular
function, and adding the constraint from the local X--ray cluster temperature
function,  joint 10\% to 30\% errors (1$\sigma$) are achievable on the
cosmological parameter pair $(\sigma_8, \OmM)$ in the flat
concordance model.
Constraints from a 2D distribution remain relatively robust
to uncertainties in possible cluster gas evolution for the case
of Planck.  Alternatively, we
examine constraints on cluster gas physics when assuming priors on the
cosmological parameters (e.g., from cosmic microwave background
anisotropies and SNIa data), finding 
a poor ability to constrain gas evolution with the 2--dimensional catalog.
From just the SZ counts and angular correlation function we obtain, 
however, a constraint on the product between the
present--day cluster gas mass fraction and the normalization
of the mass--temperature relation, $T_*$, with a precision of 15\%.
This is particularly interesting because it would be based on a very large 
catalog and is independent of any X--ray data.

   }

   \maketitle
%

\section{Introduction}

Over the coming few years, surveys of galaxy clusters observed with
the Sunyaev--Zel'dovich (SZ) effect (Sunyaev \& Zel'dovich 1970, 1972;
Birkinshaw 1999; Carlstrom et al. 2002) will open a new observational
window onto large--scale structure formation and evolution (Barbosa et
al. 1996; Eke et al. 1996; Colafrancesco et al. 1997; Diego et al
2002; Haiman et al. 2001; Holder et al. 2001; Kneissl et al. 2001;
Weller et al. 2002; Benson et al. 2002). The advantages offered by
this window, as compared to either the X--ray or the optical, are
intrinsic to the SZ effect (Bartlett 2000).  They include the ability
to detect clusters at high redshift, due to the lack of surface
brightness dimming in the SZ, and a ``clean'' selection on cluster gas
thermal energy, a robust quantity expected to have a tight
relationship to cluster mass.  These properties are particularly
advantageous for evolutionary studies because they permit the
selection of similar mass clusters over a large range of
redshifts. The distribution of cluster abundance with redshift, for
example, is sensitive to the cosmological parameters $\sigma_8$ and
$\OmM$, and also, although less so, to $\OmL$ and the dark energy
equation--of--state (Oukbir \& Blanchard 1997; Barbosa et al. 1996;
Haiman et al. 2001). This potential is currently motivating a number
of observational efforts aimed at realizing SZ surveys with dedicated,
optimized interferometers
(AMI\footnote{http://www.mrao.cam.ac.uk/telescopes/ami/index.html},
AMiBA\footnote{http://www.asiaa.sinica.edu.tw/amiba},
SZA\footnote{http://astro.uchicago.edu/sze}), and large--format
bolometer arrays (APEX\footnote{http://bolo.berkeley.edu/apexsz},
ACT\footnote{http://www.hep.upenn.edu/$\sim$angelica/act/act.html},
BOLOCAM\footnote{http://astro.caltech.edu/$\sim$lgg/bolocam\_front.htm},
ACBAR\footnote{http://cosmology.berkeley.edu/group/swlh/acbar/},
SPT\footnote{http://astro.uchicago.edu/spt/}).  The
Planck\footnote{http://astro.estec.esa.nl/Planck/} satellite, to be
launched in 2007, will provide a full-sky catalog of
galaxy clusters detected by their SZ signal, one of the largest galaxy cluster
catalogs ever constructed, and in the more distant future one may 
look forward to an even larger catalog from a fourth generation 
CMB mission, such as the {\em Inflation Probe} proposed by 
NASA in the context of the Beyond Einstein Program\footnote{{\tt 
http://universe.nasa.gov/program/probes.html}}.


\begin{table*}

\begin{flushleft}
\begin{tabular} {|c|c|c|c|c|} \hline
Survey&Y limit ($arcmin^2$)&Coverage (sq.deg.)&Average redshift&Expected number of clusters \\ 
&&&&( for $\OmM=0.27$, $\OmL=0.73$, $h=0.72$, $\sigma_8 = 0.84$)\\ \hline

Planck & $3 \times 10^{-4}$&40000&0.3&36000 \\
SPT& $5 \times  10^{-5}$&4000&0.6&33000  \\
APEX& $2.5 \times  10^{-5}$&250&0.7&5000  \\ \hline

\end{tabular}

\end{flushleft}

\caption{The surveys that have been considered in this analysis.} \label{tab-surveys}
\end{table*}


Follow--up in other wavebands of a SZ catalog is obviously essential
for many scientific goals, for instance to constrain cosmology and
cluster evolution with the redshift distribution and X--ray properties
(e.g., Holder et al. 2001; Bartelmann \& White 2002; Diego et
al. 2002; Weller et al. 2002; Hu 2003; Majumdar \& Mohr 2003a;
Majumdar \& Mohr 2003b).  Extensive follow--up will be limited to
only small subsets of the larger SZ catalogs.  In particular, the
follow--up of the Planck all--sky catalog will represent a significant
effort.  It is therefore interesting to ask the question as to what science
can be done with a two--dimensional SZ catalog, what we refer to as
the SZ {\em photometric catalog}.

In a previous paper (Mei \& Bartlett 2003, MB03), we studied the
counts and the angular correlation function of SZ clusters to see how
these two statistics could be combined to extract cosmological
information before any subsequent follow--up.  The angular function
has been extensively studied by Diaferio et al. (2003), while
three dimensional clustering issues are elaborated by Moscardini et
al. (2002).  Specifically, we explored how joint measurements of the counts
and angular function could be used to constrain the cosmological
parameters $\sigma_8$ and $\Omega_M$, when the normalization of the
Mass--Temperature relation for clusters is known. This work focused on
the influence of various cosmological parameters and cluster gas
physics on both the counts and the angular function.  In previous work, 
Fan \& Chiueh (2001) examined constraints in the $\sigma_8$--$\Omega_M$ 
plane obtained by combining a SZ catalog with limited redshift information 
(e.g., only two redshift bins) and the local abundance of 
X--ray clusters.

We extend our study in this paper by quantifying the achievable
constraints with a standard Fisher analysis, working in the context of
the so--called concordance model ($\OmM=0.27$, $\OmL=0.73$, $h=0.72$; e.g. Spergel et al. 2003).
Two kinds of constraints are used to illustrate the use of a SZ
photometric catalog: constraints on the cosmological parameter pair
$(\sigma_8, \OmM)$ in the presence of possible cluster gas evolution,
and constraints on cluster gas physics assuming strong cosmological
priors (e.g., from cosmic microwave background anisotropies and SNIa distance
measurements).  We
furthermore examine the gain obtained by incorporating the constraint
from the local X--ray cluster temperature distribution function.  

With just the photometric catalog, we find that determinations of
$\sim 10\%$ on $\sigma_8$ and $\sim 25\%$ on $\OmM$ are possible,
assuming reasonable uncertainty on cluster gas physics.  On the other
hand, gas evolution is only poorly constrained even when adopting
strong cosmological priors.  Our present study takes as examples the
almost full--sky SZ catalog expected from Planck, (Aghanim et
al. 1997; Bartelmann 2001; Diego et al. 2002; and references therein),
and two deeper ground--based experiments, one covering 4000
square degrees (e.g., Haiman et al. 2001; Holder et al. 2001; Majumdar
\& Mohr 2003b), representative of the South Pole Telescope (SPT)
survey, and one covering 250 square degrees,
representative of the APEX survey.  
The characteristics of these surveys are summarized in
Table~\ref{tab-surveys} with flux limits quoted at a
signal--to--noise of better than three. The expected number of
clusters and mean redshift have been calculated for the case of our
fiducial model (see Section~4).

In the next section we outline our cluster model, referring for details to
MB03.  The Fisher analysis is then presented in Section 3.  Section 4
presents our main results in terms of achievable constraints on
$\sigma_8$ and $\OmM$ and on cluster gas physics.  A final discussion
follows.

\begin{figure*}
\centerline{\includegraphics[width=5cm]{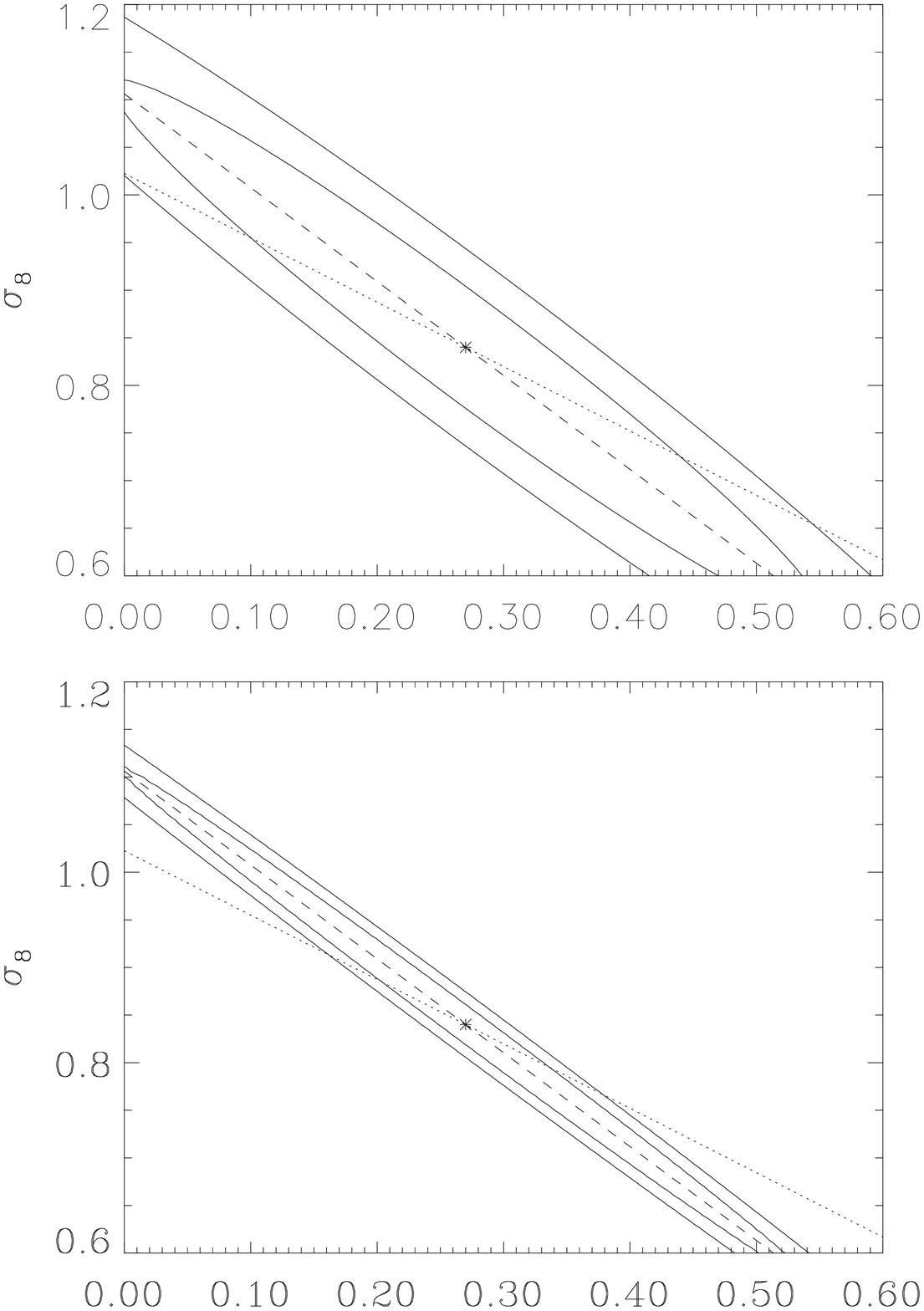}\includegraphics[width=5cm]{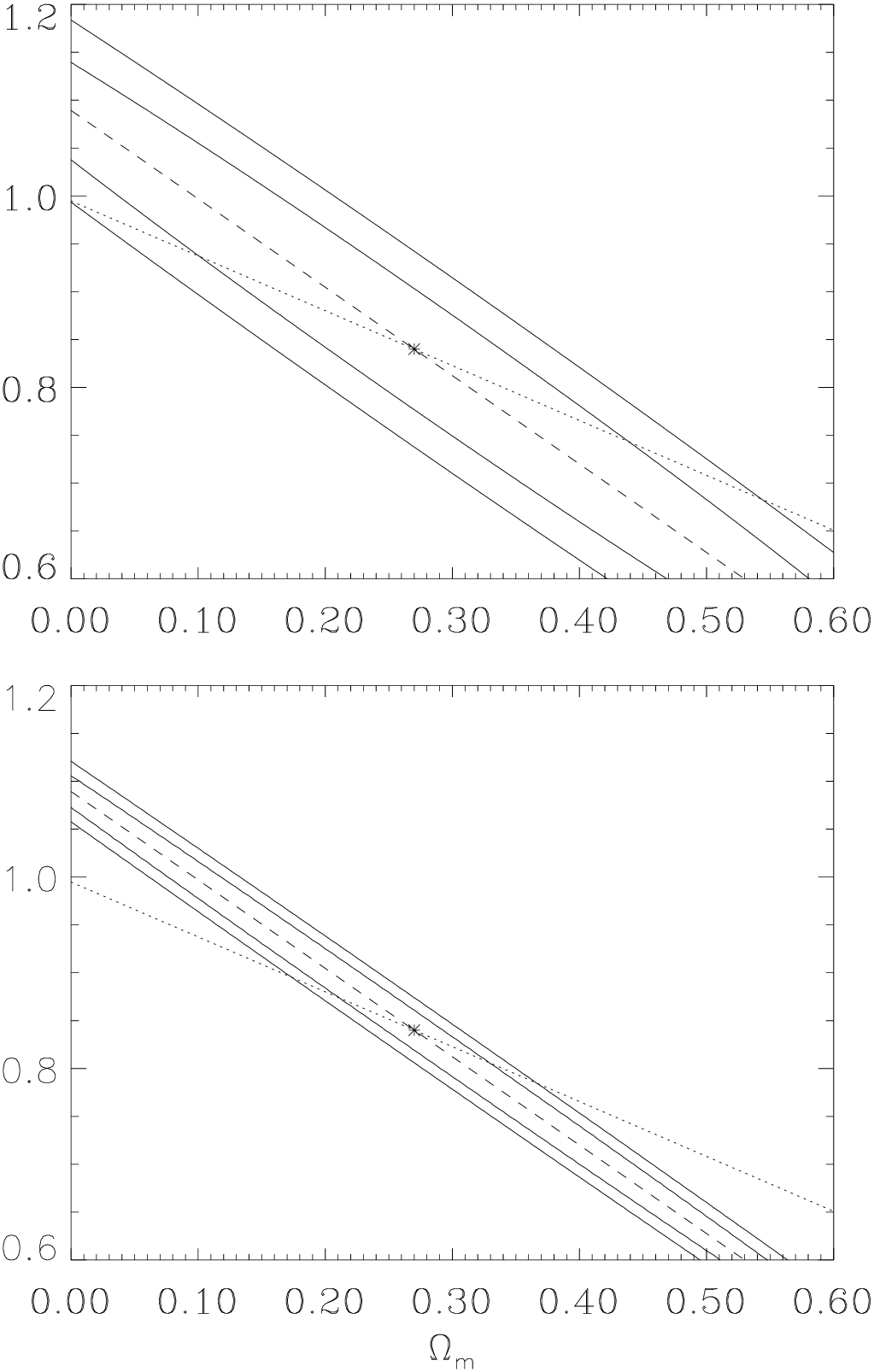}\includegraphics[width=5cm]{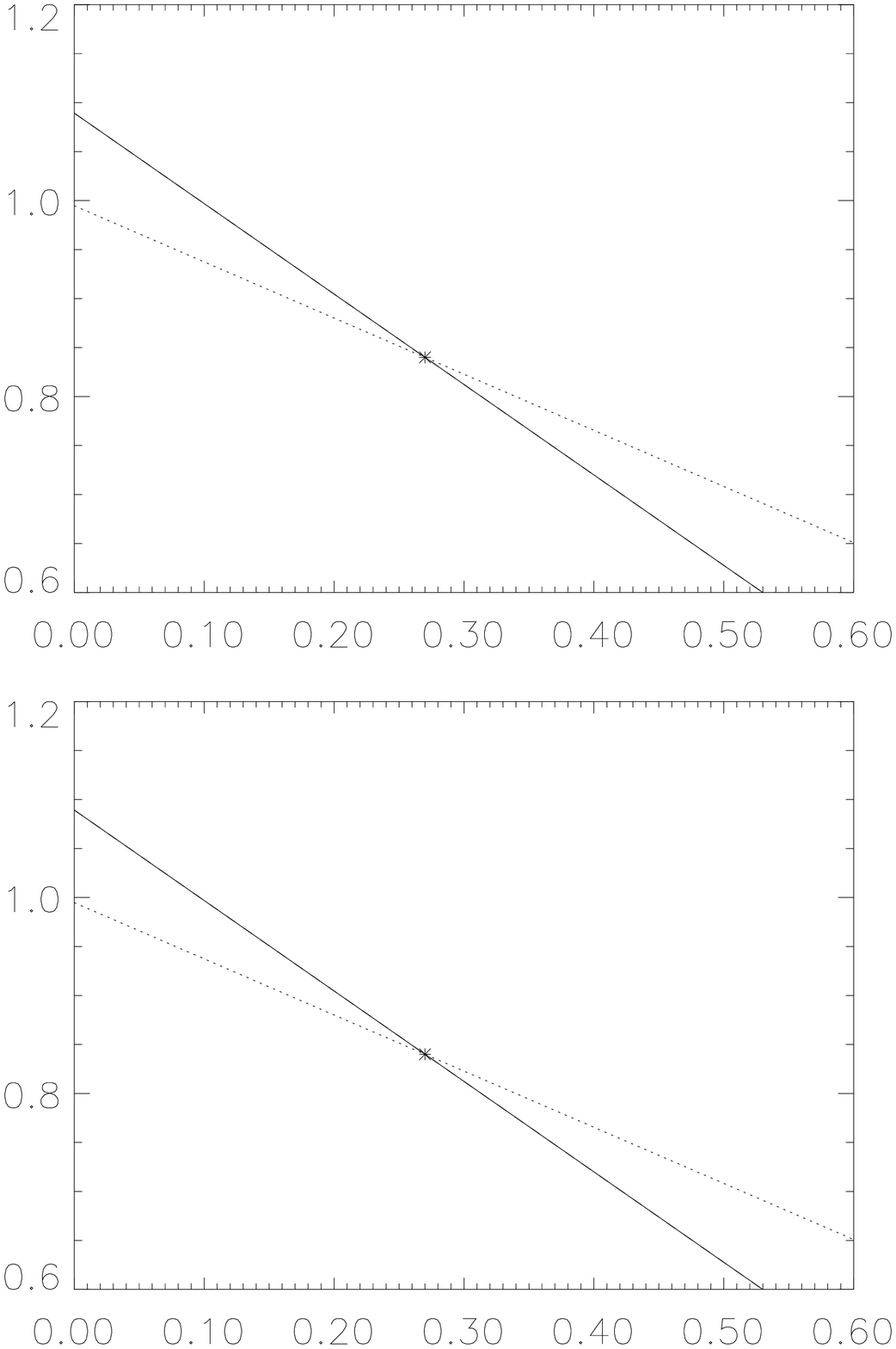}}
\caption{Joint constraints from the counts and the angular correlation 
function on $\Sei$ and $\OmM$ are shown (solid lines) at one and two
$\sigma$ (2 independent parameters), with a prior of 30\% (top) and
10\% (bottom) on $Y_*^\prime$.  Our fiducial model is shown as the
asterisk. The dashed line represents the degeneracy line from the
singular value decomposition of the count Fisher matrix.  The dotted
line represents the degeneracy line from the singular value
decomposition of the angular correlation function Fisher matrix.  From
left to right we show the constraints for a Planck--like survey, a
SPT--like survey and an APEX--like survey (see Table~1). In the case
of the APEX--like survey, joint constraints are not shown, since the
joint Fisher matrix is singular : in this case the solid line
represents the degeneracy line from the singular value decomposition
of the joint Fisher matrix; it falls directly on the degeneracy line
from the the counts Fisher matrix.}
\label{fig:twocon}
\end{figure*}

\section{SZ cluster physics}

We give only a brief summary of our main modeling ingredients, leaving
greater detail to MB03.  Properties of the cluster population reflect
the influence of halo evolution and cluster gas physics.  For the
halos, we adopt the mass function of Sheth \& Tormen (1999) and use
the expression for the linear growth factor $D(z,\OmM,\OmL)$ given by
Carroll et al. (1992).  Halo clustering is modeled with a linear bias,
$b(M,z)$, described by the analytic fit of Sheth et al. (2001).  We
use the BBKS transfer function with shape parameter fixed at
$\Gamma=0.25$ for the matter density perturbation power spectrum.
Although not critical for our final results, we include non--linear
evolution of the density field according to the prescription developed
by Peacock \& Dodds (1996).

The total SZ flux from a cluster (relative to the mean sky brightness,
i.e., the unperturbed cosmic microwave background [CMB]) is measured
by the integrated Compton y--parameter, which may be expressed in
terms of cluster quantities as
\begin{equation}\label{eq:Yscaling}
Y(M,z) = \frac{k\sigma_T}{mc^2}\frac{N_{\rm e}T}{\Da^2(z)}
       \propto \frac{\fgas(M,z) T(M,z) M}{\Da^2(z)} 
\end{equation}
where $k$ is the Boltzmann constant, $m$ is the electron mass, and
$N_{\rm e}$ is the total number of electrons in the cluster.  In this
expression, $\fgas(M,z)$ is the cluster gas mass fraction, $T(M,z)$ is
the mean {\em particle weighted} gas temperature, $M$ is the total
virial mass and $\Da(z)$ is the angular diameter distance in a
homogeneous background.  Notice that $Y$ is directly proportional to
the gas thermal energy, and we would expect this to have a tight
relationship to cluster mass and redshift.  This is indeed borne out
by hydrodynamical simulations (da Silva et al 2004), confirming the
idea that SZ selection is ``clean'' and robust.

The gas mass fraction and temperature are in general functions of
cluster mass and redshift, and we have taken care to write them 
as such.  A number of simple scaling relations
may be obtained by assuming that clusters form a self similar
population, as would be expected if non--gravitational forces
are sub--dominant (Kaiser 1986; Bryan \& Norman 1998).  Such scaling arguments
lead one to expect
\begin{equation}\label{eq:Tscaling}
T(M,z) = T_* \left(M_{15} h\right)^{2/3} \left[\Delta (z)
         E(z)^2\right]^{1/3} \left[1-2\frac{\OmL(z)}{\Delta (z)}\right]
\end{equation}
where $T_*$ is a normalization constant (expressed in keV), $M_{15}$
is the cluster total mass in units of $10^{15}$~M$_\odot$, $\Delta(z)$
is the non--linear density contrast on virialization ($\approx 178$)
and $h~\equiv~\Ho~/~100$~km/s/Mpc.  The quantity 
$E^2(z)~=~[\OmL +(1-\OmM-\OmL)(1+z)^2 + \OmM(1+z)^3]$ 
(the dimensionless Hubble parameter) with the definitions 
$\OmM(z)\equiv~\OmM(1+z)^3/E^2(z)$, $\OmL(z)\equiv~\OmL/E^2(z)$; 
notice that $\Omega_M$ and $\Omega_\Lambda$ written without an 
explicit redshift dependence will indicate present--day values 
($z=0$).  The gas mass fraction $\fgas(M,z)$ is, on the other hand, 
constant in the simplest self--similar model, independent of cluster 
mass and redshift (e.g., Arnaud et al. 2002).

Putting all this together, we express the relation between cluster SZ
flux and mass and redshift as
\begin{equation}\label{eq:Ygenscaling}
Y(M,z) = Y_{15}(z)M_{15}^{5/3+\alpha}(1+z)^\gamma
\end{equation}
where $Y_{15}(z)$ incorporates the various constants and redshift
dependence of the self--similar model.  The exponents $\alpha$ and
$\gamma$ describe any deviations from pure self--similarity, in other
words gas evolution, such that the self--similar model is defined by
$\alpha=\gamma=0$.  In their cooling hydrodynamical simulations, da
Silva et al. (2004) actually find very little deviation from
self--similarity, even down to very low masses: $\alpha\approx
0.1$ and $\gamma\approx 0$.  The explicit expression for $Y_{15}(z)$ is
\begin{eqnarray}\label{eq:Ynorm}
\nonumber
Y_{15}(z) = & \left(7.4\times 10^{-5}h^{7/6}\;{\rm arcmin}^2\right)  
        \left(\frac{T_*}{{\rm keV}}\right) 
        \left(\frac{\fgas}{0.07h^{-3/2}}\right) \times \nonumber \\
        &\left(\frac{\Delta (z) E (z)^2}{178}\right)^{1/3}
        \left[1-2\frac{\OmL(z)}{\Delta (z)}\right]
        \frac{1}{\da^2(z)}  \nonumber \\
	& \equiv  Y_*\left(\frac{\Delta (z) E(z)^2}{178}\right)^{1/3}
        \left[1-2\frac{\OmL(z)}{\Delta (z)}\right]
        \frac{1}{\da^2(z)} \nonumber \\
	&\equiv \left(1.06\times 10^{-3}h^{8/3}\;{\rm arcmin}^2\right)
          Y_*^\prime \times  \nonumber\\
         &\left(\frac{\Delta (z) E(z)^2}{178}\right)^{1/3}
        \left[1-2\frac{\OmL(z)}{\Delta (z)}\right]
        \frac{1}{\da^2(z)} \nonumber \\
\end{eqnarray}
where $\Da\equiv \Ho^{-1}\da$. 
In the following, we use $Y_*^\prime~\equiv~\fgas~T_*$ to indicate
our normalization of this relation.  A certain amount of
uncertainty remains in this constant due to uncertainties in both
$T_*$ and $\fgas$.  We will explore the effects of this uncertainty as
well as of gas evolution by treating $\alpha$, $\gamma$ and $Y_*^\prime$ as
free parameters constrained within observational limits.

\section{Fisher matrix method}

To calculate confidence regions we use the Fisher matrix method
(e.g., Eisenstein et al. 1999, Holder et al 2001), which is faster than
Monte Carlo simulations and gives accurate enough results when the
likelihood distribution is close to Gaussian.  This is the case for
the source counts that will number in the thousands (see Table 1).
Since the angular correlations of SZ clusters are
small on the scales of interest to us (e.g., tens of arcmins) and for
the relevant survey depths (Diaferio et al. 2003, MB03), the
statistical measurement error on the angular correlation function
$w(\theta)$ can be modeled by the Poissonian variance in the number of
random pairs $\npair$ at separation $\theta$ in a ring $\Dtheta$ 
(e.g., Peebles 1980;
Landy \& Szalay 1993). As discussed in MB03, this quantity is
determined by the cluster counts and equals $\npair \approx (1/2)
N\times \langle n\rangle = N \pi\theta \Dtheta\Sigma$, $N$ being the
total number of clusters in the catalog, and $\Sigma$ the surface density 
of clusters at the flux limit (cluster counts at the flux limit).  
In this case, and when the
angular correlation function is measured by optimal estimators (Landy
\& Szalay 1993), $w(\theta)$ can be treated as a Gaussian random
variable with variance equal to its Poissonian value.

The Fisher matrix is defined as

\begin{eqnarray}
F_{ij} \equiv  -\langle \frac{\partial^2 \ln \mathcal L}
{\partial p_i \partial p_j}\rangle = \langle\frac{\partial X}{\partial p_i} 
\frac{\partial  X}{\partial p_j} \frac{1}{\sigma^2_{X}}\rangle
\end{eqnarray}
where $\mathcal L$ is the likelihood for the physical variable $X$ and
the $p_i$ are the parameters we wish to constrain; the angled brackets
indicate an ensemble average over all possible data realizations.  In
our case $X$ will be respectively equal to the total number of
clusters at the survey flux limit, $N$, and $w(\theta=30$~arcmins$)$
(see MB03 for this choice of separation); their respective errors are
$\sigma^2_{N} = N$ and $\sigma^2_{w(\theta)} = \frac{1}{\npair}$. We
take the inverse of $F_{ij}$ to model the best covariance matrix
$C_{ij}$ for the considered parameters, and we use Gaussian statistics
(like $\chi^2$) to obtain one and two $\sigma$ confidence regions. We
sum the two Fisher matrices to obtain the joint confidence limits from
the combined measurement of both the counts and the angular
correlation function.  When incorporating the constraints from the
X--ray temperature function, we add the three Fisher matrices.  We
have used singular value decomposition (Press et al. 1992) for every
Fisher matrix to control singularities due to degeneracy among
parameters.

\begin{figure*}
\centerline{\includegraphics[width=5cm]{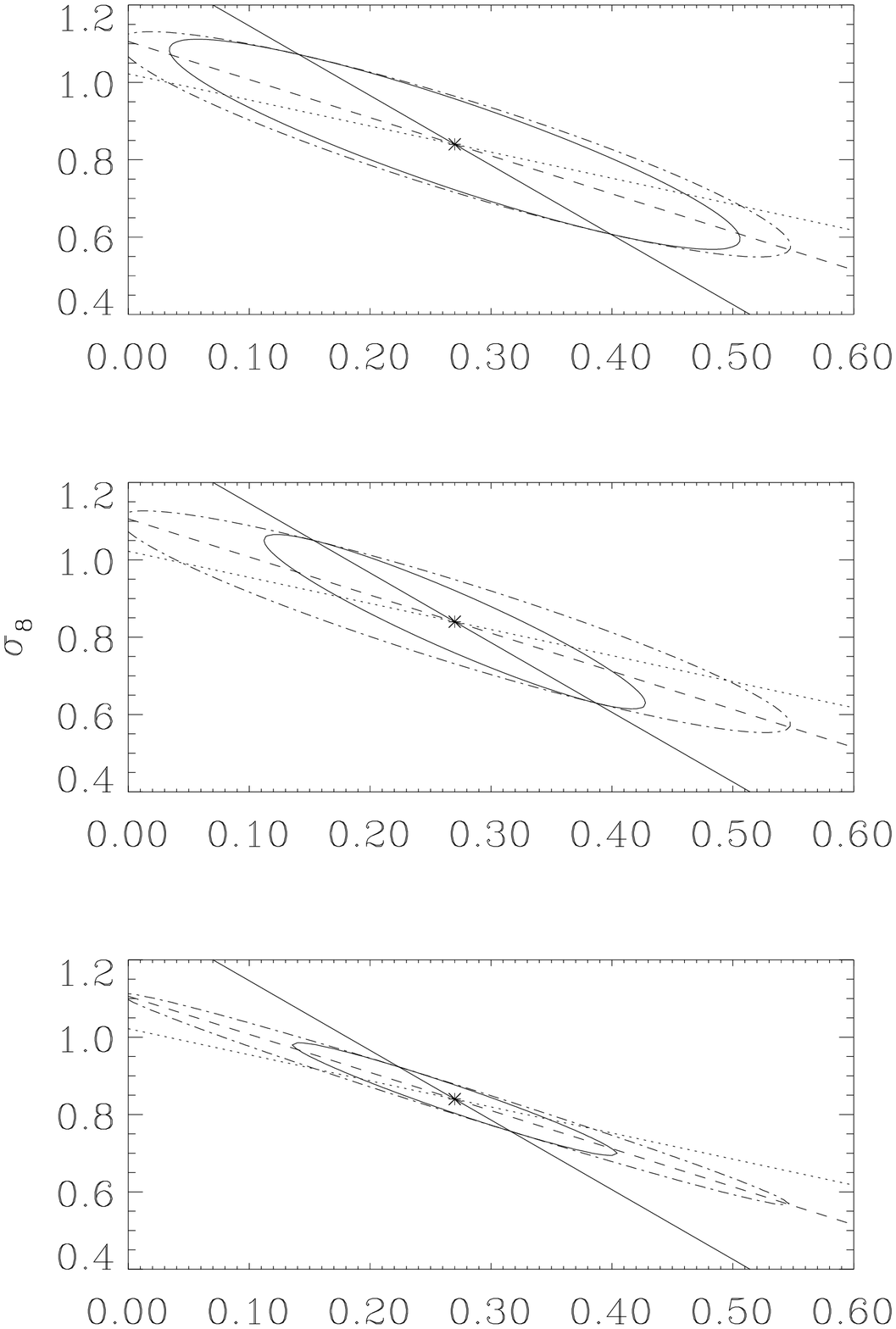}\includegraphics[width=5cm]{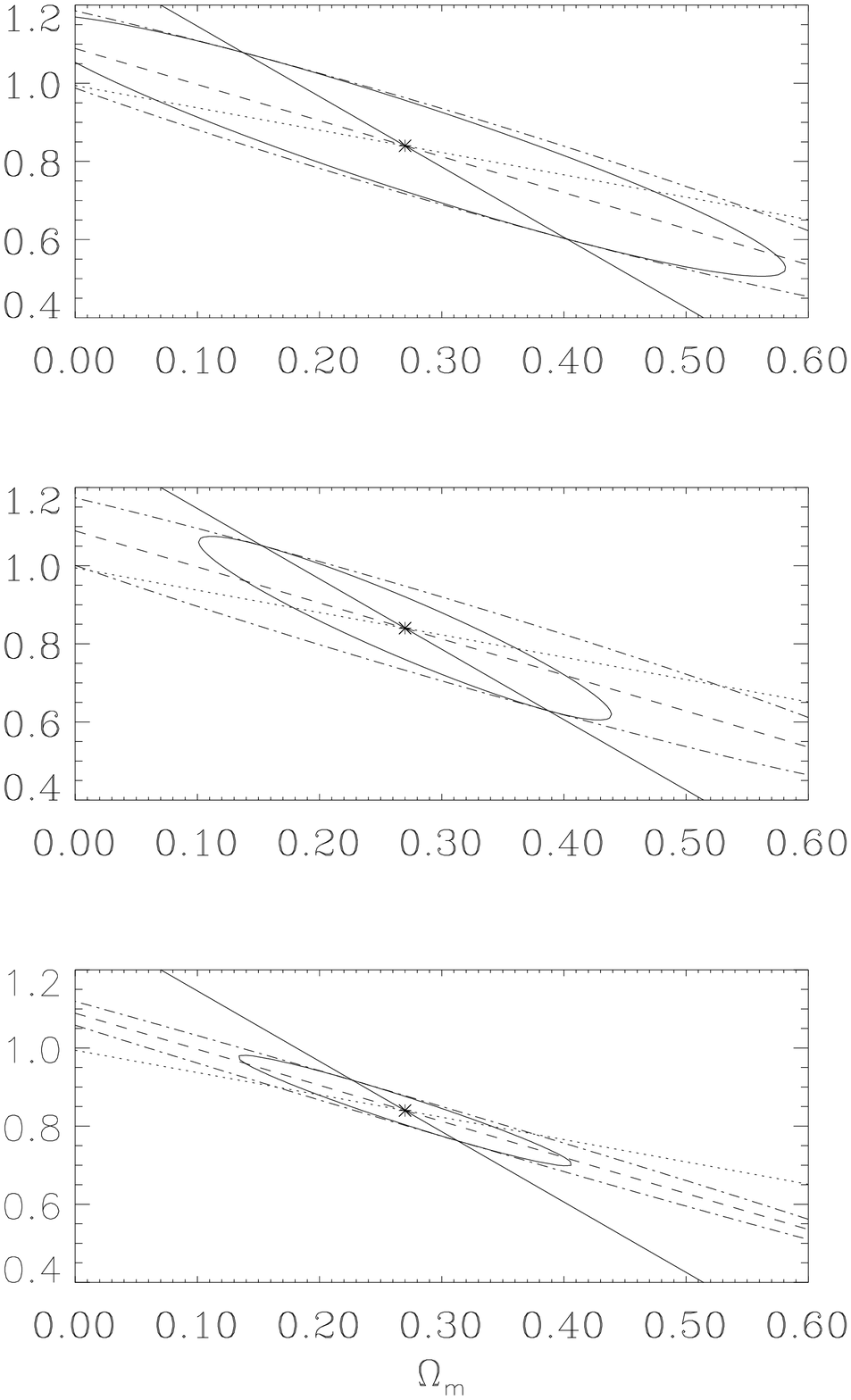}\includegraphics[width=5cm]{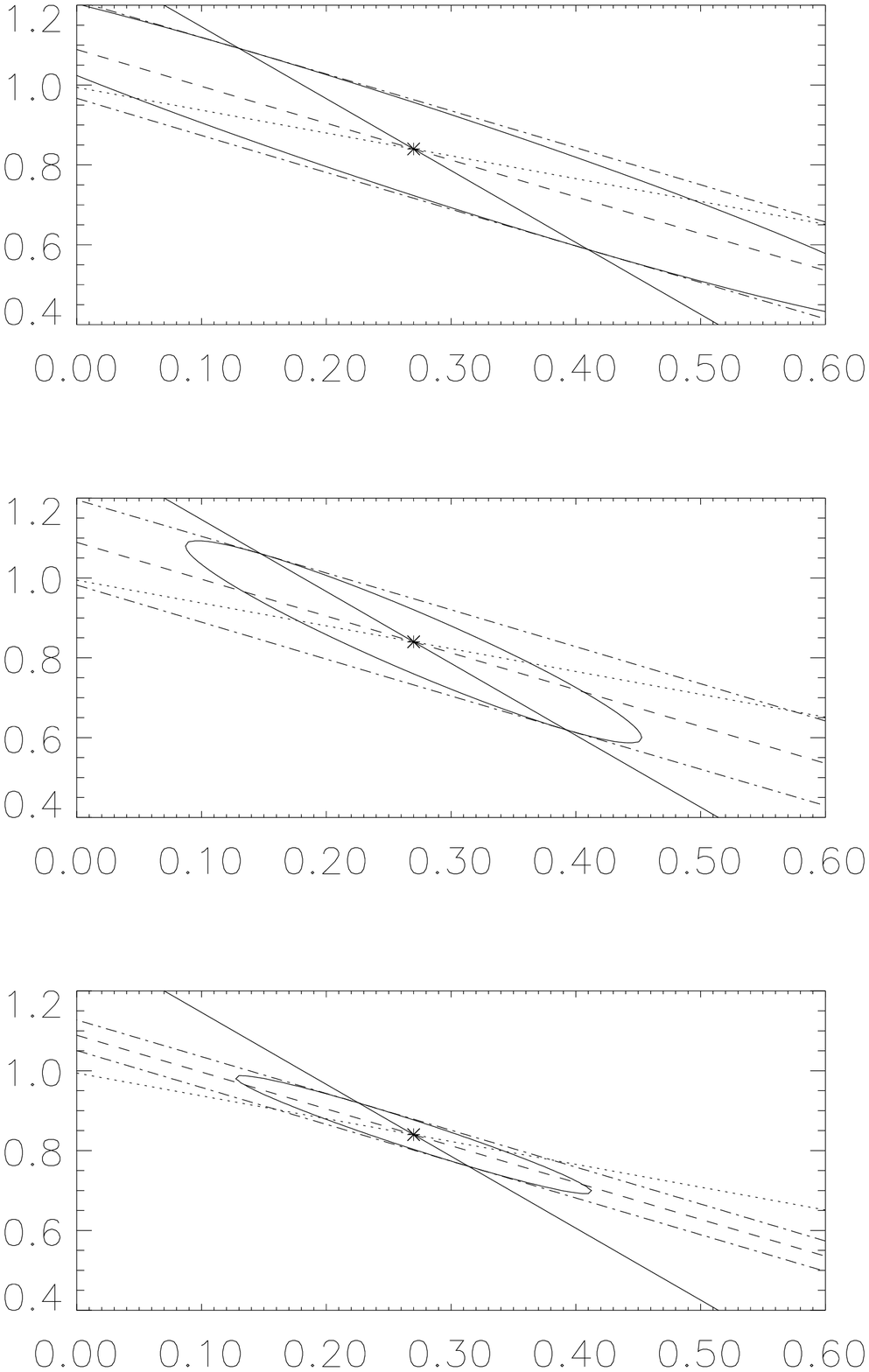}}
\caption{Constraints on $\Sei$ and $\OmM$ from a joint analysis of the 
counts, the angular function and the local X--ray temperature function
(constraint from Pierpaoli et al. 2003) are shown at one $\sigma$
(continuous ellipse). The constraints from the joint counts and
angular correlation function are shown by the dashed--dotted
ellipse. The dashed line represents the degeneracy line from the
singular value decomposition of the counts Fisher matrix.  The dotted
line represents the degeneracy line from the singular value
decomposition of the angular correlation function Fisher matrix. The
continuous line crossing the contours represents the degeneracy line
from the singular value decomposition of the Fisher matrix for the
constraints from the local X--ray temperature function. In the top
panel, priors of 30\% on $T_*$ and 50\% on $\fgas$ are assumed; in the
middle panel, priors of 10\% on $T_*$ and 50\% on $\fgas$, and in the
bottom panel, priors of 10\% on $T_*$ and 15\% on $\fgas$ are assumed.
From left to right, we show the constraints for Planck--like,
SPT--like and APEX--like surveys. These constraints are summarized in
Table~\ref{tab-fig2}.}
\label{fig:togpier}
\end{figure*}


\begin{table*}

\begin{flushleft}
\begin{tabular} {|c|c|c|c|c|c|} \hline
Survey&$T_*$ Prior Unc. (\%)&$\fgas$ Prior Unc. (\%)&$\sigma_{\Sei}$ (\%)&$\sigma_{\OmM}$(\%)&$\sigma_{\fgas}$(\%)\\ \hline

Planck & 30 &50&20&60&30\\
&10 &50&20&40&30\\
&10 &15&10&30&10\\ \hline

SPT & 30 &50&30&80&35\\
&10 &50&20&40&30\\
&10 &15&10&30&10\\ \hline

APEX & 30 &50&35&100&35\\
&10 &50&20&45&35\\
&10 &15&10&35&10\\ \hline

\end{tabular}

\end{flushleft}

\caption{One sigma constraints on $\Sei$ and $\OmM$ from a joint analysis of the 
counts, the angular function and the local X--ray temperature
function. For each survey, the prior uncertainties on $T_*$ and
$\fgas$, and the expected final constraints on $\Sei$, $\OmM$, and
$\fgas$ are given. From the joint analysis we derive constraints
$\Sei$, $\OmM$, but also gain precision on $\fgas$. This table
summarizes the results of Figure~\ref{fig:togpier}.}\label{tab-fig2}
\end{table*}


\section{Results}

We consider two types of study to illustrate the uses of a photometric
SZ catalog.  In the first case, we are interested in constraining
cosmological parameters, such as $\Sei$ and $\OmM$ (for the flat
concordance model with $h=0.72$), with just the two--dimensional
catalog and in the presence of a certain amount of unknown cluster gas
physics.  As a second illustration, we assume the cosmological
parameters are given and study the achievable constraints on the
cluster gas physics.  This might represent a study in which it is
assumed that the cosmology is already fixed by other observations,
such as measurements of CMB anisotropies and the SNIa Hubble diagram.
We examine the three types of survey listed in
Table~\ref{tab-surveys}.

\subsection{Constraints on $\Sei$ and $\OmM$}

To get a feeling for the complementarity of the total counts and the angular
function, we first examine a self--similar model with
$\alpha=\gamma=0$, but with varying $Y_*^\prime$.  The effects of gas
evolution (i.e., deviations from self--similarity with $\alpha, \gamma
\ne 0$) are studied in the second subsection.

\subsubsection{Constraints in the absence of gas evolution}

In Figure \ref{fig:twocon} we show constraints in the $\Sei$--$\OmM$
plane, marginalized over the free parameter $Y_*^\prime=\fgas T_*$.
Our fiducial model adopts $Y_*^\prime=0.17$~keV -- for example, $\fgas
= 0.07 h^{-1.5}$ (Mohr et al. 1999, see also Grego et al. 2002) and
$T_* = 1.5$~keV (Pierpaoli et al. 2003) -- and a flat cosmological
model with $\Sei = 0.84$ and $\OmM = 0.27$ (Spergel et al. 2003). 
 Current values for $T_*$,
whether from simulations or observational estimates, range from 1~keV
to about 2~keV (Pierpaoli et al. 2003; Muanwong et al. 2002; Huterer
$\&$ White 2002; Finoguenov et al. 2001; Xu et al. 2001; Nevalainen et
al. 2000; Horner et al. 1999), although we may hope that it will be
much better known by the time large SZ catalogs become available, say
with more thorough lensing studies.  We therefore take as a
representative possibility an uncertainty of 30\% on $T_*$ and 10\% on
$\fgas$ (e.g., Mohr et al. 1999; Grego et al. 2002), which leads to an
overall prior on $Y_*^\prime$ on the order of 30\% (top panel).  A
hopeful case would be to reach an overall uncertainty of 10\% (bottom
panel) on $Y_*^\prime$.

The dashed line represents the degeneracy line from the singular value
decomposition of the total counts Fisher matrix.  The dotted line represents
the degeneracy line from the singular value decomposition of the
angular correlation function Fisher matrix. The joint constraints 
from the total
counts and angular correlation function are
shown by the continuous contours at one and two sigma, and our fiducial
model is indicated by the asterisk.  Each catalog statistic is
individually highly degenerate, but as discussed at length in MB03,
the respective dependence of the two measures on $\Sei$ and $\OmM$
differ, thereby lifting the degeneracies in the case of the Planck survey.

From left to right we show the constraints for a Planck--like, a
SPT--like and an APEX--like survey. 
In the case of an APEX--like
survey, joint constraints are not shown because the joint Fisher
matrix is singular: in this case the solid line represents the
degeneracy line from the singular value decomposition of the joint
Fisher matrix. It lies directly on the degeneracy line from the
singular value decomposition of the counts Fisher matrix.


\begin{figure*}
\centerline{\includegraphics[width=5cm]{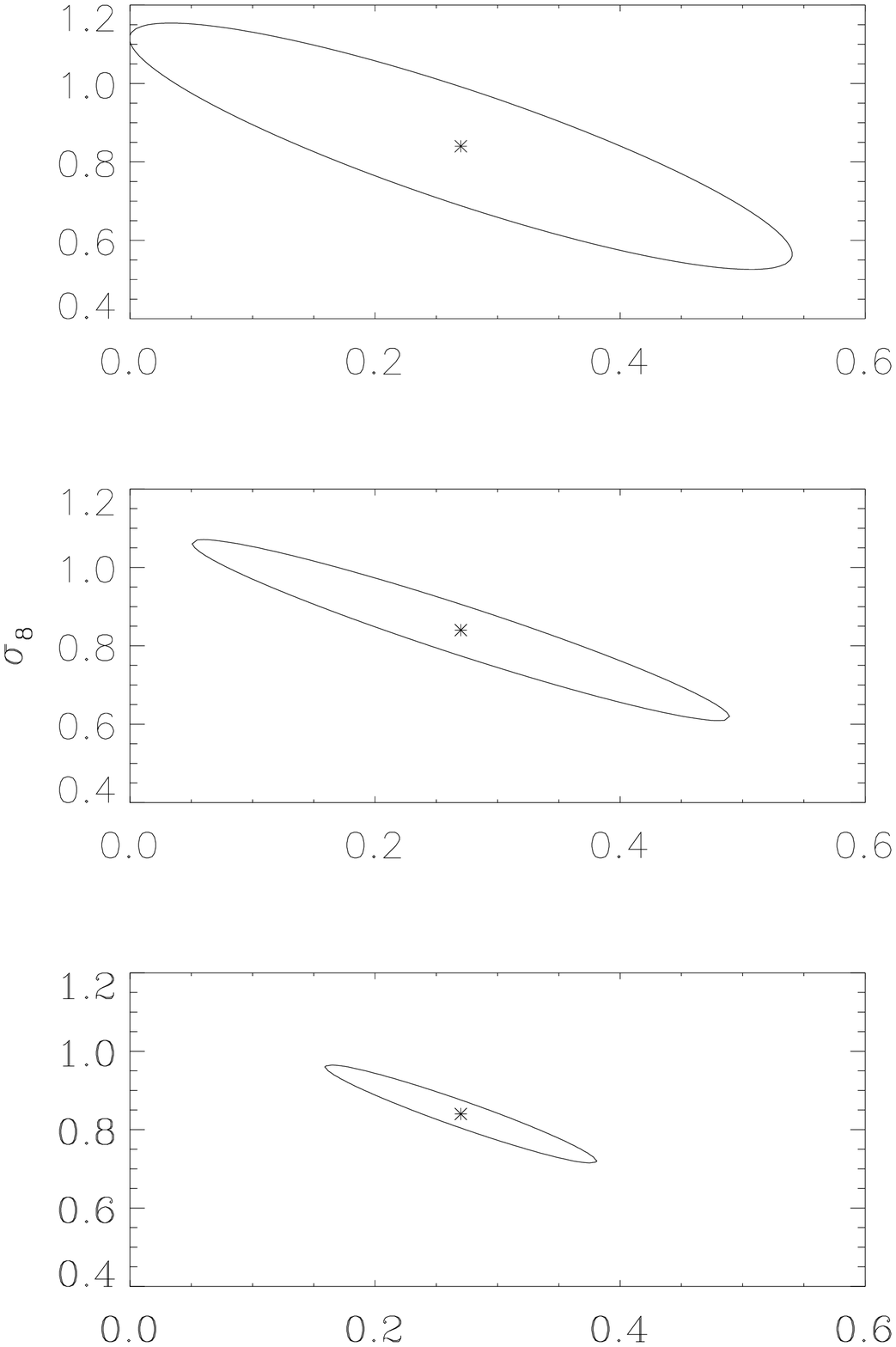}\includegraphics[width=5cm]{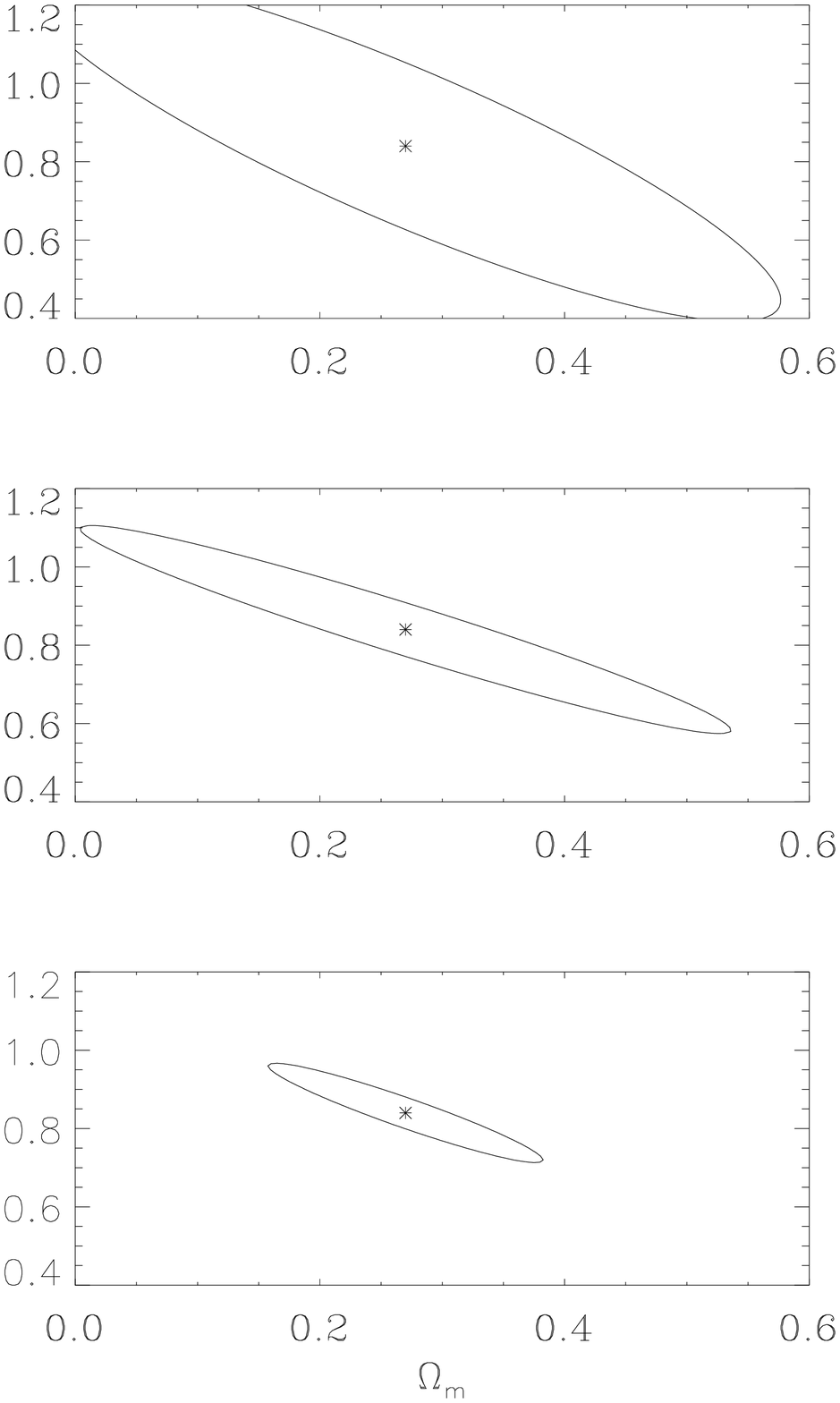}\includegraphics[width=5cm]{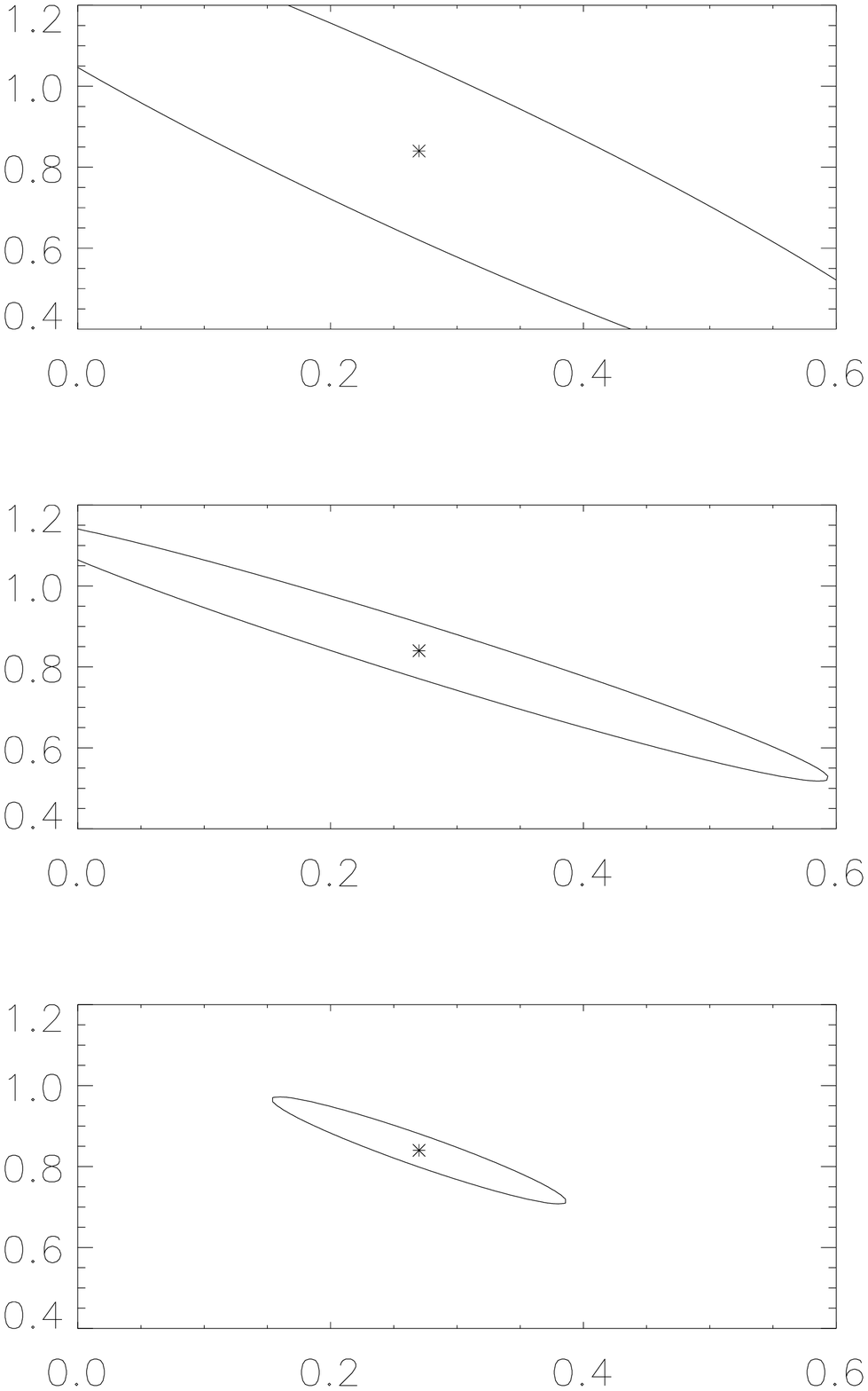}}
\caption{Constraints on $\Sei$ and $\OmM$ from a joint analysis of the 
counts, the angular function and the local X--ray temperature function
in the presence of gas evolution are shown at one $\sigma$.  In the
top panel, priors of 30\% on $Y_*^\prime$ and $\pm 1$ on $\alpha$ and
$\gamma$ have been assumed. In the center panel, the priors on
$\alpha$ and $\gamma$ are dropped to $\pm 0.1$, with the same prior on
$Y_*^\prime$. In the bottom panel, priors are taken as 15\% on
$Y_*^\prime$ and $\pm 0.1$ on $\alpha$ and $\gamma$. 
 From left to right, we show the
constraints for Planck--like, SPT--like and APEX--like surveys.  These
constraints are summarized in Table~\ref{tab-fig3}.}
\label{fig:omsig2}
\end{figure*}


\begin{table*}

\begin{flushleft}
\begin{tabular} {|c|c|c|c|c|c|c|c|c|c|} \hline
Survey& $Y_*^\prime$ Prior Unc. (\%)&$\alpha$ and $\gamma$ Prior Unc.&$\sigma_{\Sei}$ (\%)&$\sigma_{\OmM}$(\%)\\ \hline

Planck & 30 &1&30 &70\\
&30 &0.1&20&50\\
&15 &0.1&10&30\\ \hline

SPT & 30 &1&40 &80\\
&30 &0.1&20&70\\
&15 &0.1&10&30\\ \hline

APEX & 30 &1&50 &1\\
&30 &0.1&30&80\\
&15 &0.1&10&30\\ \hline

\end{tabular}

\end{flushleft}

\caption{One sigma constraints  from a joint analysis of the 
counts, the angular function and the local X--ray temperature
function. For each survey, the prior uncertainties and the final
expected constraints on the considered parameters are given. This
table summarizes the results of
Figure~\ref{fig:omsig2}.}\label{tab-fig3}
\end{table*}


Since the counts and the angular correlation function depend on  $\sigma_8, \OmM$, and $Y_*^\prime$, it is not
possible to constrain all three parameters without additional 
information.  In
Figure~\ref{fig:twocon} this information was taken as the prior on
$Y_*^\prime$.  As discussed in MB03, in fact the predicted curves for 
the counts and angular function shift
around in the $\sigma_8$--$\OmM$ plane with $Y_*^\prime$.  
Adding the local cluster
abundance constraint pins down a unique point in this plane, thereby
fixing both the cosmological parameters and the value of $Y_*^\prime$.
The local abundance of 
clusters is quantified by the X--ray temperature function and leads to a 
constraint on the parameter combination $\sigma_8\OmM^{0.6}\approx 
0.6T_*^{-0.8}$ (Pierpaoli et al. 2003, and references therein).  To apply 
this constraint, we need some prior information on $T_*$, for example from 
lensing or detailed X--ray studies.  With such a prior, a joint analysis 
of the counts, angular function and local cluster abundance yields 
constraints on $\sigma_8$, $\OmM$ and $Y_*^\prime$; the latter then also 
implies a constraint on $\fgas$.

A joint analysis on $\Sei$ and $\OmM$ from the counts, the angular
function and the X--ray temperature function constraint, as taken from
Pierpaoli et al. (2003), is shown Figure~\ref{fig:togpier}.  From left
to right we show the constraints for Planck--like, SPT--like and
APEX--like surveys.  In the top panel, priors of 30\% on $T_*$ and
50\% on $\fgas$ are assumed. In the middle panel, priors of 10\% on
$T_*$ and 50\% on $\fgas$, and the bottom panel, priors of 10\% on
$T_*$ and 15\% on $\fgas$ are assumed.  Addition of the local
abundance constraint tends to close the ends of the error ellipses.
These results are summarized in Table~\ref{tab-fig2}.  We assume that 
we already have some prior information on $\fgas$, corresponding to the 
actual observational situation.
The fact that the joint analysis leads to an independent 
constraint on $\fgas$ is demonstrated by the gain in precision on 
this parameter shown in the last column of the Table.
We conclude therefore that 10\%--30\% precision is possible on
the cosmological parameters  $\Sei$ and  $\OmM$ with a photometric
SZ catalog and we emphasize that these constraints will be independent and
complementary to those from Supernovae~Ia and CMB measurements.


\begin{figure*}
\centerline{\includegraphics[width=5cm]{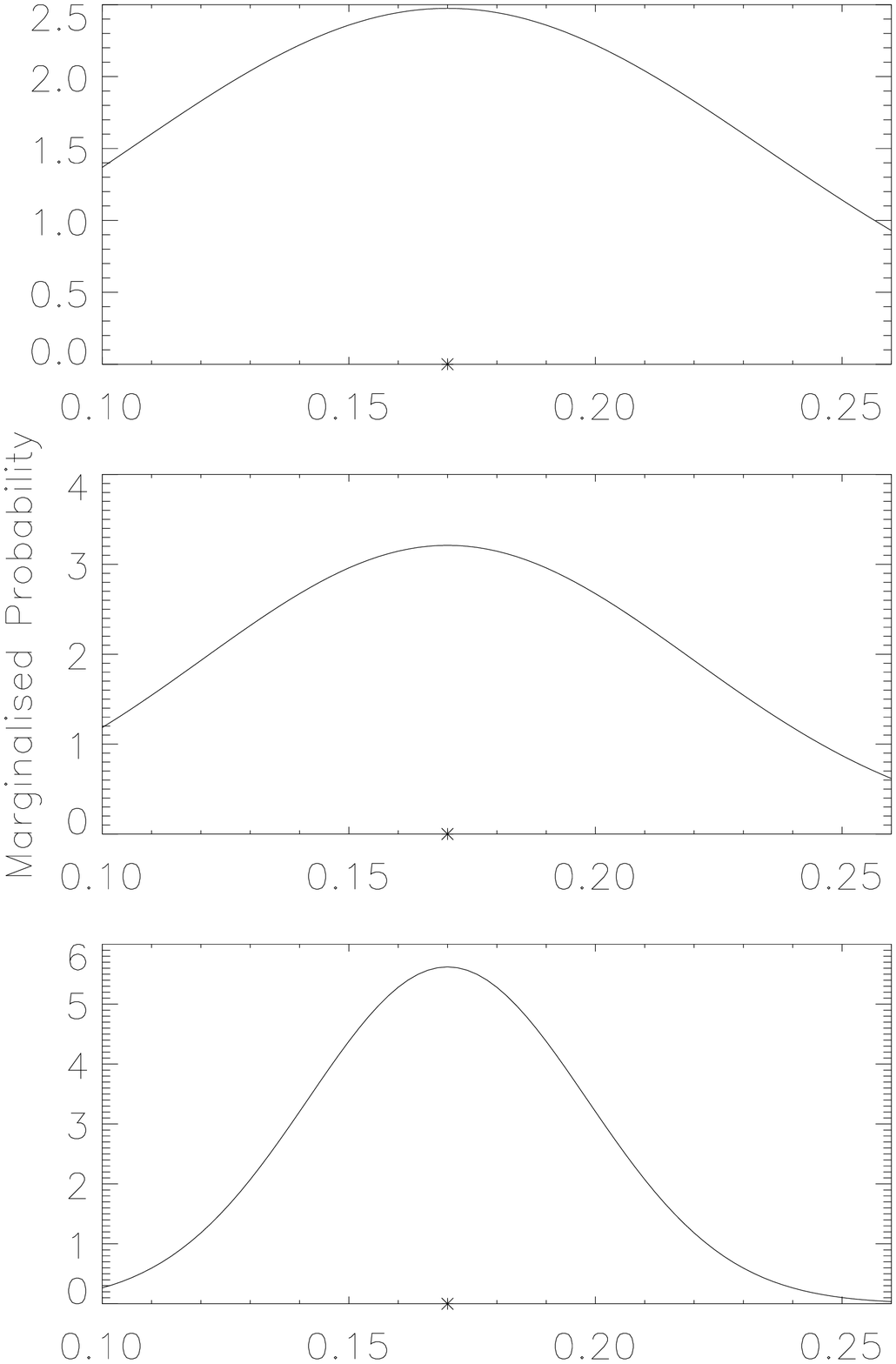}\includegraphics[width=5cm]{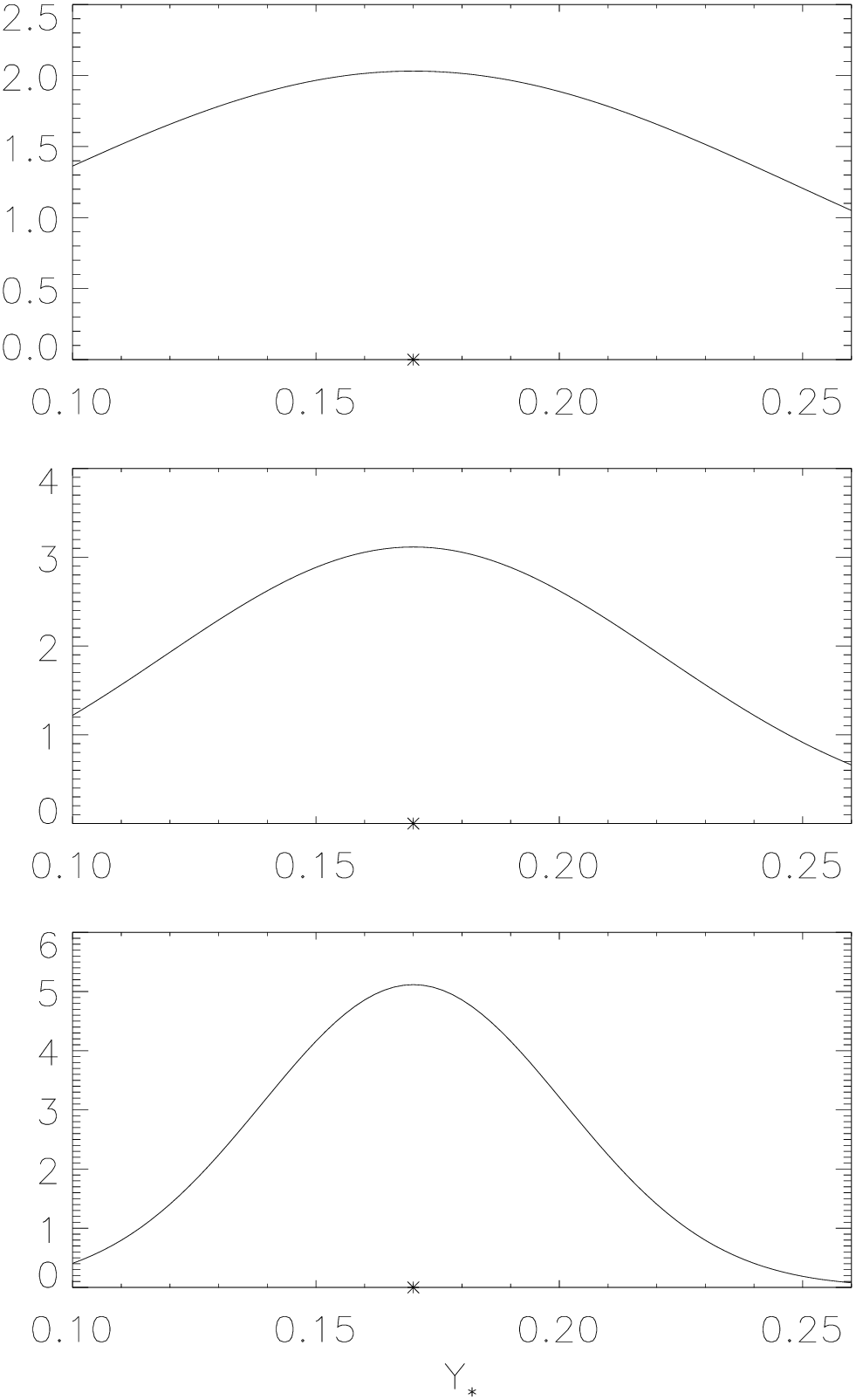}\includegraphics[width=5cm]{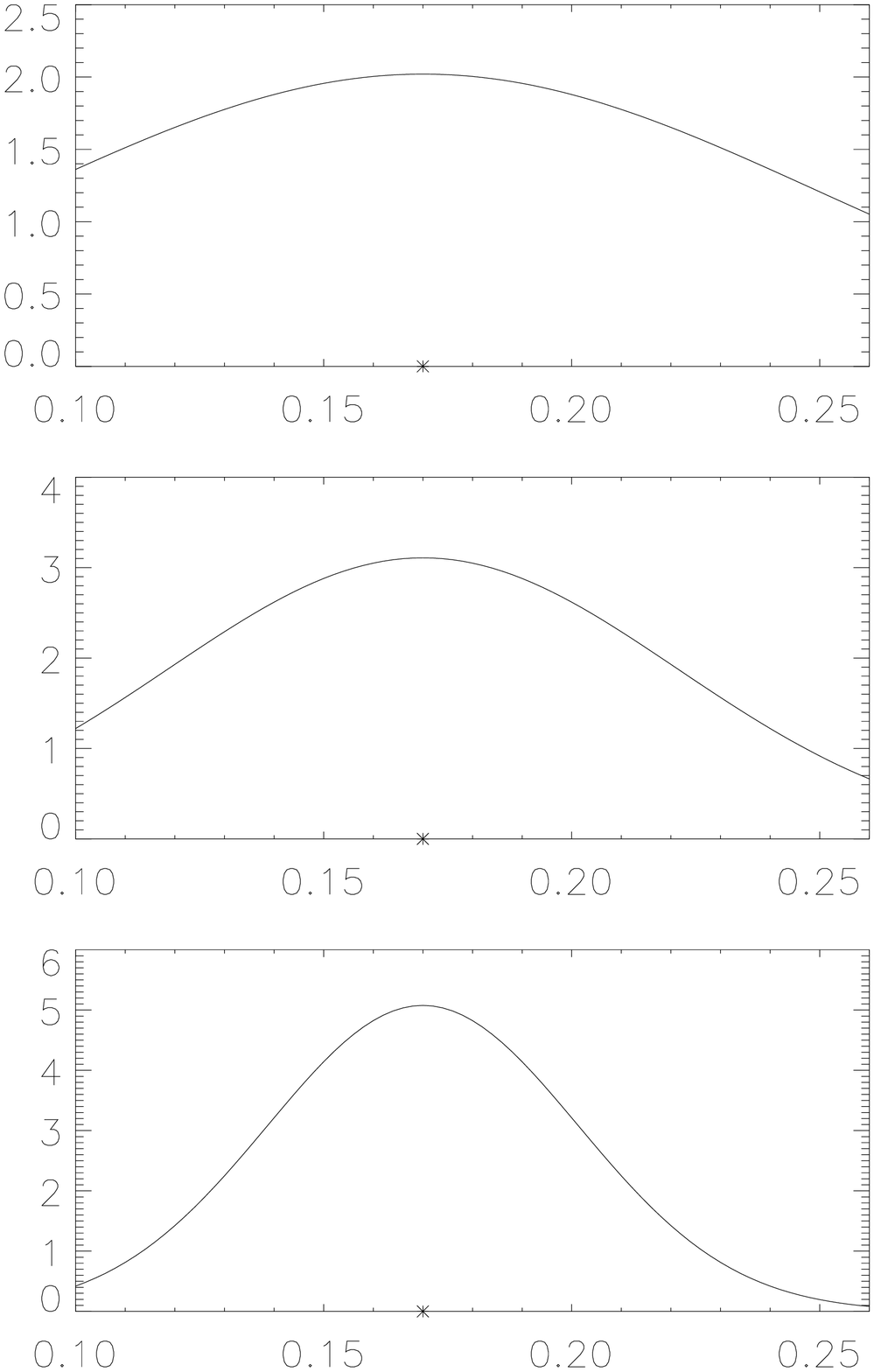}}
\caption{Constraints on $Y_*^\prime$ from a joint analysis of the 
counts and the angular function 
when the cosmological parameters are known. In the top panel, priors
are taken as 10\% on $\Sei$ and $\OmM$, and $\pm 1$ for $\alpha$ and
$\gamma$. In the center panel, the prior on $\alpha$ and $\gamma$ are
dropped to $\pm 0.1$.  In the bottom panel, priors are 5\% on $\Sei$
and $\OmM$, and $\pm 0.1$ for $\alpha$ and $\gamma$.  From left to
right, we show the constraints for Planck--like, SPT--like and
APEX--like surveys.  These constraints are summarized in
Table~\ref{tab-fb}.}
\label{fig:fb}
\end{figure*}


\begin{table*}

\begin{flushleft}
\begin{tabular} {|c|c|c|c|c|c|c|c|c|} \hline
Survey& $\alpha$ and $\gamma$ Prior Unc. &$\Sei$ and $\OmM$ Prior Unc. (\%)&$\sigma_{Y_*^\prime}$\\ \hline

Planck & 1 &10&0.5\\
& 0.1 &10&0.3\\
& 0.1 &5&0.16\\ \hline

SPT & 1 &10&0.6\\
& 0.1 &10&0.3\\
& 0.1 &5&0.18\\ \hline

APEX &  1 &10&0.6\\
& 0.1 &10&0.3\\
& 0.1 &5&0.18\\ \hline

\end{tabular}

\end{flushleft}

\caption{One sigma constraints on $Y_*^\prime$ from a joint analysis of the 
counts and the angular function 
when the cosmological parameters are known. This table
corresponds to Figure~\ref{fig:fb}.}\label{tab-fb}
\end{table*}


\subsubsection{Gas evolution}

We now examine the effects of deviations from gas self--similarity by
varying $\alpha$ and $\gamma$ in addition to $Y_*^\prime$
(Eq. \ref{eq:Ygenscaling}).  At present, very little can be said
observationally concerning the exponents $\alpha$ and $\gamma$. For a 
discussion from an observational point of view see for example
Ettori et al. (2003), and references therein.  In
their simulations, da Silva et al. (2004) find only small deviations
from self--similarity, characterized by $\alpha\approx 0.1$ and
$\gamma\approx 0$, down to very small cluster masses (well below
$10^{14}~M_\odot$) and in the presence of gas cooling.  This
illustrates the expected robustness of the SZ flux: as a measure of
the total gas energy, determined largely by gravitational collapse,
$Y$ remains rather insensitive to the details of the gas physics.  The
non--zero value of $\alpha$ reflects the change in gas thermal energy
due to cooling, which will clearly have a greater effect on the lower
mass systems.  The majority of clusters in the Planck catalog will be
rich systems (due to the high flux limit) that are relatively
insensitive to heating/cooling mechanisms. Based on these
considerations, we examine cosmological constraints with priors on
$\alpha$, $\gamma$ of $\pm 0.1$ and $\pm 1$, the latter most likely
representing an extreme case.

Figure~\ref{fig:omsig2} shows the constraints on $\Sei$ and $\OmM$
when including these new free parameters.  Our fiducial model is once
again $Y_*^\prime=0.17$~keV ($\fgas =0.07~\times~h^{-1.5}$, $h=0.72$,
$T_* =1.5$~keV) and $\alpha=0$ and $\gamma =0$. In the top panel, we
assume priors of 30\% on $Y_*^\prime$ (30\% on $T_*$) and $\pm 1$ on
the exponents $\alpha$ and $\gamma$.  These latter drop to $\pm 0.1$
for $\alpha$ and $\gamma$ in the center panel, with the same prior on
$Y_*^\prime$; and in the bottom panel, priors are taken as 15\% on
$Y_*^\prime$ (10\% on $T_*$) and $\pm 0.1$ on $\alpha$ and $\gamma$.
The priors and the
constraints are summarized in Table~\ref{tab-fig3}.  
Better knowledge
of $Y_*^\prime$ obviously improves the cosmological constraints, as
seen in going to the bottom panel of the figure.

Comparison of the top and middle panels of Figure~\ref{fig:omsig2}
shows that the constraints are affected by uncertain gas evolution
($\alpha$ and $\gamma$).  We note, however, that the Planck catalog is
less sensitive to this effect, because it primarily includes
massive clusters out to redshifts of only order unity.

\subsection{Constraints on cluster physics}

We now suppose that $\Sei$ and $\OmM$ are given and study the gas
physics parameters $Y_*^\prime$, $\alpha$ and $\gamma$.  Once again,
we adopt the concordance model with $\Sei = 0.84$, $\OmM = 0.27$
(Spergel et al. 2003) and values of $T_* = 1.5$, $\fgas=0.07
h^{-1.5}$, $\alpha =0$ and $\gamma =0$ for our fiducial model. 

Once  $\Sei$ and $\OmM$ are given, constraints on $Y_*^\prime$ can be obtained
with just SZ measurements, independently of any other external data,
 using the counts and the angular correlation function. 
Adding 
the constraints on $T_*$ from the local X--ray temperature function will
place a constraint on $f_{gas}$ and {\it vice versa} -- when external constraints
on  $f_{gas}$ are available, having the SZ derived constraints on $Y_*^\prime$
permits us to constraint  $T_*$ independently of X--ray data.

The one--dimensional likelihood function for $Y_*^\prime$ is given in Figure
\ref{fig:fb} for different priors on cosmological parameters and
$\alpha$ and $\gamma$. Since $Y_*^\prime$ is a linear combination of
$T_*$ and $f_{gas}$, constraints on $Y_*^\prime$ translate into
constraints on one of these two parameters once the other one is known. For
example, in the best case of a limit of $\approx$~15\% on $Y_*^\prime$
and a 10\% prior on $T_*$ (resp. $\fgas$), $\fgas$ (resp. $T_*$) will
be constrained to 20\%; if the prior on $T_*$ (resp. $\fgas$) is known at 5\%,
the other parameter will be constrained to 15\%.
The errors on $Y_*^\prime$ corresponding to this figure are summarized in
Table~\ref{tab-fb}.

 This kind of analysis is, on the other hand, unable to constraints the evolution parameters
$\alpha$ and $\gamma$; the $\alpha$--$\gamma$ plane is highly degenerate.

\section{Discussion \& conclusion}

Using a Fisher analysis, we have quantified the constraints achievable
with a SZ photometric catalog before any subsequent follow--up to
obtain redshifts. Our analysis has been restricted to flat cosmologies
centered on the concordance model with $\OmM=0.27$, $\sigma_8=0.84$
and $h=0.72$. The local abundance of X--ray clusters, as measured by
the present--day X--ray temperature distribution function, adds
additional information that can be usefully combined with the SZ
counts and angular function.  With prior information on $T_*$, all
three parameters $(\sigma_8,\OmM, Y_*^\prime)$ may be constrained,
which also yields a constraint on the cluster gas mass fraction
$\fgas$.  This determination of $\fgas$ would be truly representative
of the cluster population, as it is an average over a potentially very
large number of objects.  Constraints on the order of 10\% to 30\%
(around the concordance values) are obtained on the cosmological
parameters $(\sigma_8,\OmM)$ with both an all--sky
survey to $Y\sim 10^{-4}$~arcmin$^2$ and a deep ground--based survey
to $Y\sim few \times 10^{-5}$~arcmin$^2$ (see Figure~\ref{fig:togpier} and Figure~\ref{fig:omsig2}).  
To achieve these results, one must have external
information equivalent to a 10\% prior on the value of the
normalization of the $T-M$ relation ($T_*$).

These general results are not hugely affected by non--standard (i.e.,
non self--similar) gas evolution, in particular in the case of Planck.  
The corollary is that we are unable to turn
the argument around in the sense that even if the cosmological
parameters are taken as fixed, very little restriction is placed on
gas evolution.  On the other hand, if the cosmological parameters are
known, we are able, by constraining the normalization of the $Y(M,z)$
relation, to constrain the present day ($z=0$) gas mass fraction
$\fgas$ to about 20\% (with a prior of 10\% on the normalisation of
the mass/temperature relation $T_*$); or, vice versa, the normalization
of the mass/temperature relation $T_*$ to about 20\% (with a prior of
10\% on $\fgas$) .  Once again, this represents a measurement over a
very large number of clusters.

In conclusion, an angular SZ catalog in which both the counts and
angular correlation function are measured can provide useful
cosmological constraints, permitting an immediate return on a SZ
survey before subsequent follow--up observations.

\begin{acknowledgements}

S. Mei thanks Matthias Bartelmann, Frank Bertoldi and Saleem Zaroubi
for useful discussions, and acknowledges support from the European
Space Agency External Fellowship program.  Some of this work was
performed at the Lawrence Berkeley National Laboratory and at the 
University of California, Berkeley, thanks to funding from the France
Berkeley Fund (grant ``Precision Cosmology from CMB analysis''). 
\end{acknowledgements}

\end{document}